\documentstyle[preprint,aps]{revtex}

\begin{document}
\title{A Search for the Fourth SM Family Fermions and $E_{6}$ Quarks at $\mu
^{+}\mu ^{-}$ Colliders}
\author{A. K. \c{C}ift\c{c}i$^{a}$, R. \c{C}ift\c{c}i$^{b}$, S. Sultansoy$^{b,c}$}
\address{$^{a}$ Physics Dept., Faculty of Sciences, Ankara University, 06100\\
Tandogan, Ankara, Turkey\\
$^{b}$ Physics Dept., Faculty of Sciences and Arts, Gazi University, 06500\\
Teknikokullar, Ankara, Turkey\\
$^{c}$ Institute of Physics, Academy of Sciences, H. Cavid Ave., Baku,\\
Azerbaijan}
\maketitle

\begin{abstract}
The potential of $\mu ^{+}\mu ^{-}$ colliders to investigate the fourth SM
family fermions predicted by flavour democracy has been analyzed. It is
shown that muon colliders are advantageous for both pair production of
fourth family fermions and resonance production of fourth family quarkonia.
Also isosinglet quarks production at $\mu ^{+}\mu ^{-}$ colliders has been
investigated.
\end{abstract}

\section{Introduction}

The mass spectrum and the mixing of fundamental fermions are the most
important unsolved problems of the particle physics. According to the
Standard Model (SM), these masses and mixings arise from the interaction
with the Higgs doublet via spontaneous symmetry breaking.

In the framework of SM, fermions with the same quantum numbers (electric
charge, weak isospin, etc.) are indistinguishable before the symmetry
breaking. Therefore, in the fermion-Higgs interaction, the Lagrangian terms
corresponding to fermions with the same quantum numbers should come with
equal strength. As a result, one deals with singular mass matrices after the
spontaneous symmetry breaking.

According to DMM (Democratic Mass Matrix) approach \cite
{Harari,Fritzsch79,Fritzsch87,Fritzsch90,Fritzsch94} in the case of $n$ SM
families $(n-1)$ families are massless and $n^{th}$ family fermions have
masses $na\eta $ (here $a$ is the common strength of Higgs-fermion
interactions). Taking the real mass spectrum of the third family fermions
into account necessarily leads to the assumption that at least a fourth SM
family must exist \cite{Datta,Celikel,Atag} (for recent situation see Ref. 
\cite{Sultansoy}).

The existence of the fourth SM family and masses of \ the fourth SM family
quarks will be determined as a result of experiments done at LHC \cite
{ATLAS,Arik}. In our opinion, muon colliders will be advantageous for
investigation of the fourth SM family leptons and quarkonia.

\section{The Production of the Fourth SM Family Fermions \ at $\protect\mu %
^{+}\protect\mu ^{-}$ Colliders}

It is clear that direct pair production of the fourth family fermions will
be possible at future high energy colliders only, since their predicted
masses lie between $300$ $GeV$ \ and $700$ $GeV$ \cite{Celikel}. Therefore,
lepton colliders with $\sqrt{s}\geq 1.5$ $TeV$ and sufficiently high
luminosity will give the opportunity to search for all fermions from the
fourth SM family.

Linear $e^{+}e^{-}$ colliders with high energy are ones of the necessary
devices to search the fundamental ingredients of matter and interactions of
them. But the advantage of $\mu ^{+}\mu ^{-}$ colliders with respect to $%
e^{+}e^{-}$ colliders is that, they have\ more monochromatic particle beams.
For example, while the energy spread of $e^{+}e^{-}$ colliders is more than
1\%, that of $\mu ^{+}\mu ^{-}$ colliders is between 0.1\% and 0.014\%. In
addition, since the mass of muon is 207 times more than the mass of
electron, the energy uncertainty from \ effect of the opposite beam can be
ignored. Design values of $\mu ^{+}\mu ^{-}$ colliders are $\sqrt{s}=4$ $TeV$
and $L=5\times 10^{33}$ $cm^{-2}s^{-1}$ or $\sqrt{s}=30$ $TeV$ and $%
L=3\times 10^{35}$ $cm^{-2}s^{-1}$ \cite{King}.

The cross section for the process $\mu ^{+}\mu ^{-}\rightarrow f$ $\stackrel{%
\_}{f}$ \ has the form

\begin{equation}
\sigma =\frac{2\pi \alpha ^{2}}{3s}\xi \beta \left\{ Q_{f}\left( Q_{f}-2\chi
_{1}vv_{f}\right) \left( 3-\beta ^{2}\right) +\chi _{2}\left( 1+v^{2}\right) %
\left[ v_{f}^{2}\left( 3-\beta ^{2}\right) +2\beta ^{2}a_{f}^{2}\right]
\right\}
\end{equation}
where

$\chi _{1}=\frac{1}{16\sin ^{2}\theta _{W}^{2}\cos \theta _{W}}\frac{s\left(
s-M_{Z}^{2}\right) }{\left( s-M_{Z}^{2}\right) ^{2}+\Gamma _{Z}^{2}M_{Z}^{2}}
$

$\chi _{2}=\frac{1}{256\sin ^{4}\theta _{W}^{4}\cos \theta _{W}}\frac{s^{2}}{%
\left( s-M_{Z}^{2}\right) ^{2}+\Gamma _{Z}^{2}M_{Z}^{2}}$

$v=-1+4\sin ^{2}\theta _{W}$

$a_{f}=2T_{3f}$

$v_{f}=2T_{3f}-4Q_{f}\sin ^{2}\theta _{W}$

$\beta =\sqrt{1-4m_{Q}^{2}/s}$.

$T_{3}=\frac{1}{2}$ for $\nu _{4}$ and $u_{4}$, $T_{3}=-\frac{1}{2}$ for $%
l_{4}$ and $d_{4}$

$\xi =1$ for leptons, $\xi =3$ for quarks.

\bigskip The production cross section values and corresponding event numbers
(with $\sqrt{s}=4$ TeV and $L^{int}=50$ fb$^{-1}$) are given in Table I.

\bigskip

\section{The Production of the Fourth SM Family $\protect\psi _{4}\left(
^{3}S_{1}\right) $ Quarkonia at $\protect\mu ^{+}\protect\mu ^{-}$ Colliders}

Differing from t-quarks, fourth family quarks can form the quarkonia since $%
u_{4}$ and $d_{4}$ are almost degenerate and\ their decays are suppressed by
small CKM mixings \cite{Atag}.

The cross section for the formation of the fourth family quarkonium and its
decay into any $X$ state is given with the relativistic Breit-Wigner
equation 
\begin{equation}
\sigma \left( \mu ^{+}\mu ^{-}\rightarrow \left( Q\stackrel{\_}{Q}\right)
\rightarrow X\right) =\frac{12\pi \left( s/M^{2}\right) \Gamma _{\mu \mu
}\Gamma _{X}}{\left( s-M^{2}\right) ^{2}+M^{2}\Gamma ^{2}},  \label{bir}
\end{equation}
where $X$ corresponds to final state particles, $M$ is the mass of the
fourth family quarkonium, $\Gamma _{\mu \mu }$, $\Gamma _{X}$ and $\Gamma $
correspond to partial decay width to $\mu ^{+}\mu ^{-}$, $X$ state particles
and the total decay width of the fourth family quarkonium, respectively.

Since the $\mu ^{+}\mu ^{-}$ colliders has the certain energy spread, the
average cross section can be estimated from

\begin{equation}
\sigma ^{ave}=\frac{\Gamma _{tot}}{\Delta E_{coll}}\sigma ^{res}\left( \mu
^{+}\mu ^{-}\rightarrow \left( Q\stackrel{\_}{Q}\right) \right) ,
\end{equation}
where $\sigma ^{res}$ is \ the resonance value of the cross section \cite
{Barger}.

The energy spread is $\Delta E_{coll}\approx 10^{-3}\sqrt{s}$ for the $\mu
^{+}\mu ^{-}$ collider with \ $\sqrt{s}={\cal O}(TeV)$. The estimated cross
section values for $\psi _{4}\left( u_{4}\stackrel{\_}{u_{4}}\right) $ are
presented in the Table II. Correspondin values for $\psi _{4}\left( d_{4}%
\stackrel{\_}{d_{4}}\right) $ are approximately the same.

\bigskip The value of the luminosity at the resonance, used in calculations,
has been estimated as

\[
L\left( \sqrt{s_{res}}\right) =\frac{\sqrt{s_{res}}}{4TeV}L\left(
4TeV\right) . 
\]

As a result, we obtain number of events per year which are given in the last
\ column of the Table II.

In this study we consider only $\psi _{4}\left( ^{3}S_{1}\right) $ quarkonia
state. Using corresponding formulae from \cite{Barger}, we obtain decay
widths for main decay modes of $\psi _{4}\left( u_{4}\stackrel{\_}{u_{4}}%
\right) $ which are given in Table III. One can see that dominant decay
modes for $\psi _{4}$ quarkonia are $\psi _{4}\rightarrow W^{+}W^{-}$, $\psi
_{4}\rightarrow Z^{0}\gamma $ and $\psi _{4}\rightarrow \gamma H$.

\bigskip

\section{\protect\bigskip The Production of the $E_{6}$ Quarks at $\protect%
\mu ^{+}\protect\mu ^{-}$ Colliders}

Another way to explain the relation $m_{b,\tau }<<m_{t}$ is the introduction
of exotic fermions. Let us consider as an example the extension of the SM
fermion sector which is inspired by $E_{6}$ GUT model initially suggested by
F. Gursey and collaborators \cite{Gursey,GurseyS}. It is known that this
model is strongly favored in the framework of SUGRA (see Ref. \cite{Hewett}
and references therein). For illustration let us restrict ourselves by quark
sector:$\ \ \ \ \ \ \ $%
\begin{eqnarray*}
&&\left( 
\begin{array}{c}
u_{L}^{0} \\ 
d_{L}^{0}
\end{array}
\right) ,u_{R}^{0},d_{R}^{0};\ \ \ \ \ \ \ \ \left( 
\begin{array}{c}
c_{L}^{0} \\ 
s_{L}^{0}
\end{array}
\right) ,c_{R}^{0},s_{R}^{0};\ \ \ \ \ \ \ \ \ \ \left( 
\begin{array}{c}
t_{L}^{0} \\ 
b_{L}^{0}
\end{array}
\right) ,t_{R}^{0},b_{R}^{0} \\
&&\ D_{1L}^{0},D_{1R}^{0};\ \ \ \ \ \ \ \ \ \ \ \ \ \ \ \ \
D_{2L}^{0},D_{2R}^{0};\ \ \ \ \ \ \ \ \ \ \ \ \ \ \ \ \ \ \ \ \ \ \
D_{3L}^{0},D_{3R}^{0}.
\end{eqnarray*}

According to Flavor Democracy the down quarks' mass matrix has the form:$\ \
\ \ \ \ \ \ \ \ \ \ \ \ \ \ \ \ \ \ \ \ \ \ \ \ \ $%
\[
M^{0}=\left( 
\begin{array}{cccccc}
a\eta & a\eta & a\eta & a\eta & a\eta & a\eta \\ 
a\eta & a\eta & a\eta & a\eta & a\eta & a\eta \\ 
a\eta & a\eta & a\eta & a\eta & a\eta & a\eta \\ 
M & M & M & M & M & M \\ 
M & M & M & M & M & M \\ 
M & M & M & M & M & M
\end{array}
\right) , 
\]
where $M$ is the scale of ''new'' physics which determines the masses of the
isosinglet quarks. As the result we obtain 5 massless quarks and the sixth
quark has the mass $3M+m_{t}$.

$E_{6}$ quarks can be produced at $\mu ^{+}\mu ^{-}$ colliders. To estimate
the production cross sections, one can use equation (1) with following minor
changes. For the isosinglet $D_{1}$ quark, $a_{f}=0$ and $v_{f}=-4Q\sin
^{2}\theta _{W}$. The result of estimations shows that the production cross
section changes between 1.91 fb and 1.86 fb (which corresponds to events
rates between 95 and 93 per year) for the isosinglet quark mass between 0.1
TeV and 1 TeV.

Additionally, we estimated production cross sections and events numbers per
year for isosinglet $\psi $ quarkonia. They are given in Table IV. Decay
widths for main decay modes of $\psi _{D_{1}}$ are given in Table V, where
we use $\left| V_{D_{1}t}\right| \simeq 0.01$

\ 

\section{Conclusion}

We have shown that $\mu ^{+}\mu ^{-}$ colliders with $\sqrt{s}={\cal O}(TeV)$
is a good place to investigate both fourth family fermions, quarkonia and $%
E_{6}$ isosinglet quarkonia. In this study we have concentrated on $\psi
_{4}\left( ^{3}S_{1}\right) $ state, other quarkonium states will be
considered on future study.

\begin{table}[t]
\caption{The production cross section values for the fourth SM family
fermions.}
\begin{center}
{\footnotesize 
\begin{tabular}{|c|c|c|c|c|c|c|c|c|}
\hline
$M_{4}$ & \multicolumn{2}{c|}{$\mu ^{+}\mu ^{-}\rightarrow u_{4}$ $\stackrel{%
\_}{u_{4}}$} & \multicolumn{2}{c|}{$\mu ^{+}\mu ^{-}\rightarrow d_{4}$ $%
\stackrel{\_}{d_{4}}$} & \multicolumn{2}{c|}{$\mu ^{+}\mu ^{-}\rightarrow
l_{4}^{+}$ $l_{4}^{-}$} & \multicolumn{2}{c|}{$\mu ^{+}\mu ^{-}\rightarrow
\nu _{4}$ $\stackrel{\_}{\nu _{4}}$} \\ \cline{2-9}
$(GeV)$ & $\sigma (fb)$ & $Ev./year$ & $\sigma (fb)$ & $Ev./year$ & $\sigma
(fb)$ & $Ev./year$ & $\sigma (fb)$ & $Ev./year$ \\ \hline
$300$ & $9.8$ & $490$ & $5.0$ & $250$ & $6.1$ & $305$ & $1.3$ & $65$ \\ 
\hline
$375$ & $9.7$ & $485$ & $5.0$ & $250$ & $6.1$ & $305$ & $1.3$ & $65$ \\ 
\hline
$450$ & $9.7$ & $485$ & $4.9$ & $245$ & $6.1$ & $305$ & $1.3$ & $65$ \\ 
\hline
$525$ & $9.6$ & $480$ & $4.8$ & $240$ & $6.1$ & $305$ & $1.3$ & $65$ \\ 
\hline
$675$ & $9.5$ & $475$ & $4.7$ & $235$ & $6.0$ & $300$ & $1.2$ & $60$ \\ 
\hline
$750$ & $9.4$ & $470$ & $4.6$ & $230$ & $6.0$ & $300$ & $1.2$ & $60$ \\ 
\hline
\end{tabular}
}
\end{center}
\end{table}

\begin{table}[t]
\caption{The production cross section values and event numbers per year for
the fourth SM family ($u_{4}\overline{u}_{4}$) quarkonia.}
\begin{center}
{\footnotesize 
\begin{tabular}{|c|c|c|c|c|c|}
\hline
$M_{\psi _{4}}(GeV)$ & $\sigma ^{res}\ \left( pb\right) $ & $\Gamma
_{tot}\left( \psi _{4}\right) $ $(MeV)$ & $\Delta E_{coll}$ $(GeV)$ & $%
\sigma ^{ave}$ $\left( pb\right) $ & $Ev./year$ \\ \hline
$600$ & $68.2$ & $8.3$ & $0.60$ & $0.94$ & $7100$ \\ \hline
$750$ & $19.5$ & $21.1$ & $0.75$ & $0.55$ & $5200$ \\ \hline
$900$ & $6.8$ & $46.9$ & $0.90$ & $0.35$ & $4000$ \\ \hline
$1050$ & $2.8$ & $93.3$ & $1.05$ & $0.25$ & $3300$ \\ \hline
$1200$ & $1.3$ & $170.5$ & $1.20$ & $0.18$ & $2800$ \\ \hline
$1350$ & $0.6$ & $291.6$ & $1.35$ & $0.13$ & $2200$ \\ \hline
$1500$ & $0.3$ & $472.6$ & $1.50$ & $0.09$ & $1800$ \\ \hline
\end{tabular}
}
\end{center}
\end{table}

\begin{table}[t]
\caption{Decay widths for main decay modes of $\protect\psi _{4}\left( u_{4}%
\stackrel{\_}{u_{4}}\right) $, for $m_{H}=150$ GeV.}
\begin{center}
{\footnotesize 
\begin{tabular}{|c|c|c|c|c|c|c|c|}
\hline
$M_{\psi _{4}},$ GeV & 600 & 750 & 900 & 1050 & 1200 & 1350 & 1500 \\ \hline
$\Gamma (\psi _{4}\rightarrow \ell ^{+}\ell ^{-})$, 10$^{-2}$ MeV & 1.4 & 1.6
& 1.8 & 2.0 & 2.1 & 2.3 & 2.5 \\ \hline
$\Gamma (\psi _{4}\rightarrow u\stackrel{\_}{u})$, 10$^{-2}$ MeV & 2.3 & 2.6
& 3.0 & 3.3 & 3.5 & 3.8 & 4.1 \\ \hline
$\Gamma (\psi _{4}\rightarrow d\stackrel{\_}{d})$, 10$^{-2}$ MeV & 1.0 & 1.2
& 1.3 & 1.5 & 1.6 & 1.7 & 1.8 \\ \hline
$\Gamma (\psi _{4}\rightarrow Z\gamma )$, MeV & 0.4 & 0.7 & 1.1 & 1.7 & 2.4
& 3.3 & 4.4 \\ \hline
$\Gamma (\psi _{4}\rightarrow ZZ)$, 10$^{-2}$ MeV & 4.0 & 7.6 & 12.6 & 19.2
& 27.6 & 38.0 & 50.1 \\ \hline
$\Gamma (\psi _{4}\rightarrow ZH)$, 10$^{-2}$ MeV & 4.3 & 7.9 & 13.0 & 19.6
& 28.1 & 38.5 & 51.1 \\ \hline
$\Gamma (\psi _{4}\rightarrow \gamma H)$, 10$^{-2}$ MeV & 36.1 & 66.3 & 108.5
& 164.2 & 234.8 & 321.9 & 426.8 \\ \hline
$\Gamma (\psi _{4}\rightarrow W^{+}W^{-})$, MeV & 6.8 & 18.9 & 43.6 & 88.7 & 
164.4 & 283.6 & 462.4 \\ \hline
\end{tabular}
}
\end{center}
\end{table}

\begin{table}[t]
\caption{The production cross section values and event numbers per year for
the $\protect\psi $($D_{1}\overline{D}_{1}$) quarkonia of $E_{6}$ isosinglet
quarks.}
\begin{center}
{\footnotesize 
\begin{tabular}{|c|c|c|c|c|c|}
\hline
$M_{\psi _{D_{1}}}(GeV)$ & $\sigma ^{res}\ \left( pb\right) $ & $\Gamma
_{tot}\left( \psi _{4}\right) $ $(MeV)$ & $\Delta E_{coll}$ $(GeV)$ & $%
\sigma ^{ave}$ $\left( pb\right) $ & $Ev./year$ \\ \hline
$300$ & $26.6$ & 13.8 & $0.3$ & $1.22$ & $4590$ \\ \hline
$600$ & $19.2$ & $7.1$ & $0.6$ & $0.23$ & $1700$ \\ \hline
$900$ & $1.76$ & 44.6 & $0.9$ & $0.087$ & $980$ \\ \hline
$1200$ & $0.31$ & $169.8$ & $1.2$ & $0.044$ & $660$ \\ \hline
$1500$ & $0.082$ & $478.0$ & $1.5$ & $0.026$ & $490$ \\ \hline
$1800$ & $0.028$ & $1111.7$ & $1.8$ & $0.017$ & $390$ \\ \hline
2$100$ & $0.011$ & $2270.1$ & $2.1$ & $0.012$ & $310$ \\ \hline
\end{tabular}
}
\end{center}
\end{table}

\begin{table}[t]
\caption{Decay widths for main decay modes of $\protect\psi \left( D_{1}%
\stackrel{\_}{D_{1}}\right) $, for $m_{H}=150$ GeV.}
\begin{center}
{\footnotesize 
\begin{tabular}{|c|c|c|c|c|c|c|c|}
\hline
$M_{\psi _{D_{1}}},$ GeV & 300 & 600 & 900 & 1200 & 1500 & 1800 & 2100 \\ 
\hline
$\Gamma (\psi _{4}\rightarrow \ell ^{+}\ell ^{-})$, keV & 2.3 & 3.4 & 4.3 & 
5.2 & 6.0 & 6.8 & 7.5 \\ \hline
$\Gamma (\psi _{4}\rightarrow u\stackrel{\_}{u})$, \ keV & 2.6 & 3.8 & 4.9 & 
5.9 & 6.8 & 7.7 & 8.5 \\ \hline
$\Gamma (\psi _{4}\rightarrow d\stackrel{\_}{d})$, \ keV & 0.9 & 1.2 & 1.5 & 
1.8 & 2.0 & 2.3 & 2.5 \\ \hline
$\Gamma (\psi _{4}\rightarrow ZH)$, \ keV & 4 & 27 & 82 & 177 & 321 & 524 & 
792 \\ \hline
$\Gamma (\psi _{4}\rightarrow \gamma H)$, keV & 12 & 90 & 271 & 587 & 1067 & 
1739 & 2627 \\ \hline
$\Gamma (\psi _{4}\rightarrow W^{+}W^{-})$, MeV & 0.3 & 7 & 44 & 166 & 465 & 
1082 & 2217 \\ \hline
\end{tabular}
}
\end{center}
\end{table}

\end{document}